\documentclass{IOS-Book-Article}

\usepackage{mathptmx}

\usepackage{tikz}
\usetikzlibrary{positioning,calc,arrows.meta,decorations.markings,fit,backgrounds,shapes.geometric}
\usepackage{graphicx}
\usepackage[inline]{enumitem}
\usepackage{listings}
\usepackage{booktabs}  
\usepackage{amssymb}   
\usepackage{amsmath}   
\usepackage{url}       
\usepackage{subcaption} 

\newtheorem{theorem}{Theorem}[section]

\newcommand{\sym}[1]{\text{\small\texttt{:{#1}}}}
\newcommand{\fnsym}[1]{\text{\footnotesize\texttt{:{#1}}}}
\newcommand{\facet}[1]{\sym{#1}}
\newcommand{\clabel}[1]{\sym{#1}}  
\newcommand{\framework}[1]{\text{\small{#1}}}

\begin{document}

\pagestyle{headings}
\begin{frontmatter}

\title{Parajudica: An RDF-Based Reasoner and Metamodel for Multi-Framework Context-Dependent Data Compliance Assessments}
\runningtitle{Parajudica: A Reasoner for Data Compliance Assessments}

\author[A]{\fnms{Luc} \snm{Moreau}},
\author[B]{\fnms{Alfred} \snm{Rossi}}
and
\author[C]{\fnms{Sophie} \snm{Stalla-Bourdillon}}

\runningauthor{L. Moreau, A. Rossi, S. Stalla-Bourdillon}
\address[A]{University of Sussex, Brighton, United Kingdom}
\address[B]{Immuta Research, Boston, Massachusetts, USA}
\address[C]{Brussels Privacy Hub, Vrije Universiteit Brussel, Brussels, Belgium}

\begin{abstract}
Motivated by the challenges of implementing policy-based data access control (PBAC) under multiple simultaneously applicable compliance frameworks, we present Parajudica, an open, modular, and extensible RDF/SPARQL-based rule system for evaluating context-dependent data compliance status. We demonstrate the utility of this resource and accompanying metamodel through application to existing legal frameworks and industry standards, offering insights for comparative framework analysis. Applications include compliance policy enforcement, compliance monitoring, data discovery, and risk assessment.
\end{abstract}


\bigskip
\noindent\textbf{Type:} RDF-Based Compliance Reasoner\\
\textbf{License:} CC-BY-4.0\\
\textbf{DOI:} \url{https://doi.org/10.5281/zenodo.17825090}\\
\textbf{Repository:} \url{https://github.com/alfredr/parajudica}

\end{frontmatter}

\section{Introduction}\label{sec:introduction}


Determining the permissibility of a data processing activity is an inherently complex task that often requires reasoning across multiple, overlapping policies and regulations. Such determinations are rarely dictated by the data alone; rather, they depend on contextual factors such as the roles of individuals involved, the purposes and methods of collection, jurisdictional constraints, the simultaneous availability of related data, the strength of organizational controls, assumptions about adversarial capabilities, and temporal considerations.

Determining the compliance status of the data itself is often an important first step, as access policies typically depend on these classifications~\cite{xacml,nist-sp-800-95,nist-ir-7657}. However, complicating matters, the exact same dataset containing patient information may be classified as Protected Health Information (PHI) under the U.S. Health Insurance Portability and Accountability Act (HIPAA)~\cite{hipaa-privacy} when processed by a covered entity, but as special category personal data under the General Data Protection Regulation (GDPR)~\cite{gdpr}. While data considered de-identified under either of HIPAA's regulatory standards for de-identification (safe harbor or expert determination) statutorily falls outside of HIPAA's scope, it may remain controlled as anything from anonymized data to special category personal data under the GDPR depending upon the exact details. In short, compliance status is context-dependent.

Structured data further complicates this task. A dataset containing only diagnosis, gender, and age information with a randomly assigned record identifier might be considered suitable for public release; yet if other datasets reuse the same identifier while adding additional demographic details, the released information may not be considered de-identified. The availability of identifiers, therefore, must be considered, as their presence or availability may drastically change the data compliance status.

Further complicating matters, an organization's internal compliance rules often depend on classifications made under others. For example, a policy might state that data classified as PHI under HIPAA requires encryption and audit logging, that personal data under GDPR is prohibited from cross-border transfer, or that certain combinations of domain-specific fields may be identifiable in other contexts. Such rules create interdependencies among frameworks, whereby the resulting compliance status goes beyond the disjoint union of framework determinations.

\paragraph{Terminology.}
Before presenting our approach, we define three central terms used throughout this paper. We use \emph{compliance framework} to mean organizational rules for interpreting regulatory standards (e.g., GDPR, HIPAA) and classifying data. A \emph{governance scope} represents organizational boundaries within which compliance is evaluated. The term \emph{metamodel} reflects that our approach provides structures and semantics from which specific compliance scenarios are instantiated.

\paragraph{Approach.}
To address these challenges, we formulate compliance assessment of structured data as a computational problem: propagating semantic annotations through data structures and across relationships. Data containers (databases, tables, columns) are organized hierarchically, and compliance labels propagate according to framework-specific rules: inward to contained elements, outward to containers, among peers, or across joinable relationships. Parajudica is implemented in RDF, adding to a rich tradition of established use in adjacent areas of privacy vocabularies (DPV~\cite{pandit2019dpv}), policy languages (ODRL~\cite{odrl-model-2018,odrl-vocab-2018}), and data governance initiatives~\cite{bader2020}.

Our approach is complementary to normative reasoning systems that resolve conflicts by imposing superiority relations on rules. In defeasible deontic logic (DDL), for example, the goal is to determine whether a normative statement, such as ``it is obligatory to encrypt patient data'', holds after resolving conflicts and exceptions \emph{within a single normative system}. Our problem is different. We observe that multiple internally coherent compliance interpretations (e.g., of HIPAA and GDPR) may each classify the same dataset differently. These interpretations coexist without a shared authority to adjudicate between them.

Consider two authorities with incompatible labeling schemes. If authority $A$ classifies a field as $L_1$ and authority $B$ classifies the same field as $L_2$, there is no legitimate meta-rule determining which classification prevails. Any resolution ($L_1$, $L_2$, both, or neither) would be rejected by at least one authority. Defeasible reasoning assumes such a hierarchy of defeat can be defined; multi-framework compliance does not provide one. Our metamodel instead retains both classifications ($L_1$ under $A$, $L_2$ under $B$) as parallel outputs rather than a logical inconsistency to be resolved. This is practically useful: a company operating under multiple jurisdictions may need to evaluate whether a proposed de-identification strategy satisfies each framework's requirements across the jurisdictions where it operates, despite their divergent interpretations.

Each framework in our model defines its own hierarchy of labels and propagation rules, with possible interdependencies. Within a framework, derived classifications accumulate until least fixed point. At the same time, different frameworks may adopt incompatible hierarchies. Despite possible cross-framework dependencies, resulting even in possible cycles, the results remain ``locally'' monotonic (monotonic when restricted to a single framework) but globally non-monotonic in the sense that contradictions only arise when comparing results across frameworks.

\paragraph{Limitations.}
We acknowledge that this design has limitations. Local monotonicity, for instance, prevents modeling mutually canceling conditions. Full expressiveness, however, comes at the cost of computational complexity. Rather than pursue full generality, we accept these trade-offs in favor of a well-grounded system for context-dependent classification that remains tractable. To demonstrate that practical utility remains despite these constraints, we model and recreate a series of real-world compliance challenges involving divergent interpretations and cross-framework interdependencies.

\paragraph{Contributions.}
This paper makes three contributions. First, we present Parajudica, a reasoner and metamodel for context-dependent compliance assessment of structured data that captures how the same data receives different classifications based on governance scope, available relationships, and applicable frameworks. Second, we provide ready-to-use implementations of real-world frameworks (HIPAA with Safe Harbor and Expert Determination, GDPR, EMA, and Italian DPA) that encode regulatory requirements as computational rules deriving compliance classifications through propagation. Third, we provide mathematical foundations proving polynomial-time convergence despite complex cross-framework dependencies.

\paragraph{Organization.}
The remainder of this paper is structured as follows. Section~\ref{sec:legal} identifies five compliance challenges through real-world scenarios. Section~\ref{sec:concepts} introduces our metamodel's concepts for representing data containers, compliance frameworks, and their relationships. Section~\ref{sec:modeling} demonstrates how to model specific frameworks (Base, HIPAA, GDPR, the Italian DPA's approach to anonymization under GDPR) using these concepts. Section~\ref{sec:validation} validates that our approach addresses all five challenges through a healthcare scenario. Section~\ref{sec:formalization} provides the mathematical foundations and proves convergence properties. Section~\ref{sec:implementation} describes the RDF-based implementation. Section~\ref{sec:related-conclusion} positions our work relative to existing semantic technologies and concludes with future directions.

\section{Compliance Challenges}\label{sec:legal}

This context-dependent nature of compliance assessment defeats static annotation approaches, which cannot capture the situational nuances of the broader compliance environment. To illustrate these challenges, we turn to healthcare, where organizations must reconcile multiple frameworks with conflicting requirements.

\paragraph{HIPAA.}
The Health Insurance Portability and Accountability Act (HIPAA)~\cite{hipaa-privacy} operates as a binary switch: data is either Protected Health Information (PHI), subject to all regulatory requirements, or entirely outside its scope, as in the case of de-identified data. A common interpretation extends PHI status transitively, such that any data joined to the patient record becomes part of the patient’s \emph{designated record set}~\cite{hipaa-privacy}. Yet HIPAA’s scope is context-dependent, and identical data may be regulated or not depending on its system of record and purpose. For example, employee vaccination records are excluded when maintained for workplace safety\footnote{Employee records are excluded from the definition of PHI as per 45 C.F.R. § 160.103.}, but become PHI when entered into clinical systems. Similarly, Social Security numbers must be removed from patient record extracts before they can be considered de-identified, yet their presence in employee records does not trigger HIPAA protections, even when the employer is a clinic and the employee is also a patient.

\paragraph{GDPR.}
The General Data Protection Regulation (GDPR)~\cite{gdpr} takes a different stance. Rather than closure through joins, GDPR distinguishes data \emph{identifying} individuals from data merely \emph{about} them. Removing identifiers may render data anonymous, though regulators disagree on whether anonymization is achievable~\cite{stalla_rossi_2021,stalla_2025} and it takes time for the Court of Justice of the European Union case law to mature~\cite{srb}. The European Medicines Agency (EMA), in the context of clinical trial data publication, recommends risk-based thresholds (e.g., permitting disclosure when the re-identification probability falls below 0.09)~\cite{ema_external_guidance_2016,ema_external_guidance_2025}. By contrast, the Italian Data Protection Authority (DPA), drawing on Article 29 Working Party guidance~\cite{wp29-anonymization}, rejects singling out: datasets remain personal data whenever unique identifiers enable individual distinction, regardless of actual re-identification risk~\cite{italian-thin-2023}.

Consider a clinic joining employee vaccination records with patient treatment records. Under HIPAA's transitive approach, the entire dataset becomes PHI, since employee identifiers inherit PHI status through association. Under GDPR, only vaccination information qualifies as special category health data, while employee identifiers may retain their original classification as ordinary personal data, depending on context~\cite{lindenapotheke}. This divergence underscores the need for framework-specific propagation rules. A similar pattern appears with de-identification: a dataset may be released only if the risk of re-identification is acceptably low.

One common method is \emph{k-anonymization}, which ensures that each individual record is indistinguishable from at least $k-1$ others with respect to quasi-identifier attributes. A $k$-anonymized dataset tagged only with a random identifier may be considered suitable for release under HIPAA Expert Determination (45 C.F.R. § 164.514(b)(1)), conditionally acceptable under EMA Policy 0070~\cite{ema-policy0070,ema_external_guidance_2025} with $k \geq 12$, yet always prohibited under the Italian DPA’s interpretation~\cite{italian-thin-2023}. HIPAA balances data and context risk\footnote{HIPAA thresholds reflect common practice. OCR guidance~\cite{ocr-deid-guidance} does not specify numerical standards for ``very small'' risk, but $k=5$ is a common recommendation in practice~\cite{el-emam-2009}.}, EMA imposes explicit thresholds, and the Italian DPA rejects risk-based reasoning entirely. These differences highlight the incompatibility of static compliance definitions across jurisdictions.
\begin{table}[ht]
\centering
\small
\setlength{\tabcolsep}{4pt}
\renewcommand{\arraystretch}{0.9}
\begin{tabular}{lccc}
\toprule
\textbf{$k$ Value} & \textbf{HIPAA} & \textbf{GDPR (EMA)} & \textbf{GDPR (Italian)} \\
\midrule
$k \geq 12$ & \checkmark & \checkmark & $\times$ \\
$k = 5$ & \checkmark & $\times$ & $\times$ \\
$k < 3$ & $\times$ & $\times$ & $\times$ \\
\bottomrule
\end{tabular}
\caption{Regulatory suitability of public release of $k$-anonymized datasets (with random record identifiers) under HIPAA Expert Determination, EMA Policy 0070 guidance, and Italian DPA interpretation of GDPR.}
\label{tab:regulatory-divergence}
\end{table}
\vspace{-\baselineskip}

\paragraph{Compliance Challenges.}
From the preceding discussion we distill five challenges to compliance assessment:
\begin{enumerate*}[label=(\arabic*)]
  \item \emph{context-dependent classification}, where identical data may receive different statuses across environments (e.g., vaccination records treated as HR data in one setting but PHI in clinical systems);
  \item \emph{diverging interpretations}, where the same dataset produces contradictory outcomes within one environment (e.g., joined employee–patient records treated as PHI in one framework but split into identifiers and medical data in another, with differing risk thresholds);
  \item \emph{proximity semantics}, where the availability of nearby data within a scope or operation alters compliance status (e.g., employee identifiers acquiring PHI status when linked to patient treatments);
  \item \emph{de\-identification scenarios}, which should be modeled in a way that allows reasoning from both structural assertions (e.g., removal of enumerated identifiers) and statistical assertions (e.g., $k \geq 3$, $k \geq 12$), while also accommodating frameworks that reject risk-based methods altogether;
  \item \emph{comparative evaluation}, where side-by-side analysis makes divergences explicit (e.g., Table~\ref{tab:regulatory-divergence} shows a $k=5$ dataset acceptable under one framework but prohibited under others)
\end{enumerate*}.

\section{Metamodel Description}\label{sec:concepts}

Our metamodel provides a formal foundation for reasoning about multi-framework compliance through five concepts:
\begin{enumerate*}[label=(\roman*)]
\item \emph{containers} representing data storage hierarchically,
\item \emph{frameworks} encoding regulatory regimes like GDPR or HIPAA,
\item \emph{labels} classifying data properties,
\item \emph{assertions} tracking which containers have which labels, and
\item \emph{scopes} defining compliance environment boundaries.
\end{enumerate*}

\paragraph{Data Containers and Relations.}
Containers represent data storage locations (databases, tables, columns, fields) organized through an acyclic containment hierarchy. The containment relation is irreflexive and acyclic, forming a forest structure where each container has at most one parent. Containers sharing a parent are siblings. Orthogonal to this hierarchy, explicit joinability declarations mark which containers can exchange data during operations, enabling proper tracking of compliance properties across transformations.

\paragraph{Compliance Labels, Facets, and Frameworks.}
Frameworks represent specific compliance interpretations such as GDPR, HIPAA, or internal policies. Each framework introduces a vocabulary of labels and rules that govern their behavior through subclass hierarchies, conditional equivalences, and propagation patterns across container hierarchies and joinable relations. Labels are atomic classification units that belong to facets, which organize cross-cutting concerns such as identifiability, controlled categories, or risk levels. Labels may also be parameterized to capture quantitative requirements, for example $k$-anonymity values or retention periods. Together, these elements encode regulatory logic in computational form.

\paragraph{Framework Inheritance.}
Frameworks extend others, inheriting rules and label hierarchies. When EMA extends GDPR, it inherits GDPR's label taxonomy (e.g., SpecialCategoryData $\subseteq$ PersonalData) and propagation rules, while adding stricter $k$-anonymity thresholds. This enables jurisdictional variations without duplicating base logic. Child frameworks may override inherited rules.

\paragraph{Assertions and Scopes.}
Governance scopes represent compliance environments providing context for assertions. Containers exist independently of scopes and may appear in multiple environments simultaneously with different compliance labels in each. Initially, containers enter scopes with ground assertions (fundamental properties like containing names or medical codes); framework rules derive additional labels during inference. Compliance assertions track framework-assigned labels to containers within scopes. Containment assertions declare labels within containers, with parent containers transitively containing descendant labels.

\paragraph{Rule Types.}
All rules reduce to implications: if a container meets certain conditions, it must also receive a new label. The simplest case is label-to-label: if $A$ then $B$. More generally, conditions alone may yield a label without prerequisites. Propagation rules replicate labels across structure (parent–child, child–parent, siblings, or joinable containers). Equivalence rules bind labels to apply together, often bridging frameworks. Though semantically equivalent to implications, we distinguish subclass rules for clarity of intent; it follows that labels may belong to multiple distinct subclasses. Collectively, rules match assertions, check conditions, and derive or propagate new labels, ensuring compliance properties extend consistently through hierarchies and contexts.

\paragraph{Release Operations.}
Releases enable data movement across scope boundaries, transferring only ground assertions while leaving context-dependent determinations within their original environments, preserving scope-specific compliance evaluations while allowing controlled data sharing.

\section{Modeling}\label{sec:modeling}
We instantiate our metamodel with concrete compliance frameworks to demonstrate both expressiveness and practical utility. The \framework{Base} framework provides foundational taxonomies and classification logic, which subsequent frameworks extend and specialize. \framework{HIPAA} and \framework{GDPR} illustrate distinct regulatory approaches outlined in Section~\ref{sec:legal}, while the \framework{Italian DPA}'s interpretation shows how jurisdictional rules layer atop base regulations.

\paragraph{Base.}
The \framework{Base} framework establishes a six-facet taxonomy: \facet{Subject} categorizes entities (\clabel{Individual}, \clabel{Organization}); \facet{Identifier} derives classifications from orthogonal properties; \facet{Kind} organizes semantic types with multi-inheritance; \facet{Domain} specifies operational context (\clabel{Healthcare}, \clabel{Financial}); \facet{Statistical} enables quantitative analysis, including SPARQL queries that compute \clabel{KAnonymityAnalysis} assertions; and \facet{Control} provides namespace for regulatory frameworks. \facet{Type}, \facet{Kind}, and \facet{Domain} propagate inward to child containers. Figure~\ref{fig:base-framework-structure} illustrates the taxonomy and an identifier derivation rule.

\makeatletter
\newcount\dirtree@lvl
\newcount\dirtree@plvl
\newcount\dirtree@clvl
\def\dirtree@growth{%
  \ifnum\tikznumberofcurrentchild=1\relax
    \global\advance\dirtree@plvl by 1
    \expandafter\xdef\csname dirtree@p@\the\dirtree@plvl\endcsname{\the\dirtree@lvl}
  \fi
  \global\advance\dirtree@lvl by 1\relax
  \dirtree@clvl=\dirtree@lvl
  \advance\dirtree@clvl by -\csname dirtree@p@\the\dirtree@plvl\endcsname
  \pgf@xa=0pt\relax 
  \pgf@ya=-0.35cm\relax
  \pgf@ya=\dirtree@clvl\pgf@ya
  \pgftransformshift{\pgfqpoint{\the\pgf@xa}{\the\pgf@ya}}%
  \ifnum\tikznumberofcurrentchild=\tikznumberofchildren
    \global\advance\dirtree@plvl by -1
  \fi
}
\tikzset{
  dirtree/.style={
    growth function=\dirtree@growth,
    every node/.style={anchor=north,font=\footnotesize\ttfamily},
    every child node/.style={anchor=west,font=\footnotesize\ttfamily},
    edge from parent path={([xshift=10pt]\tikzparentnode.south west) |- (\tikzchildnode\tikzchildanchor)}
  }
}
\makeatother

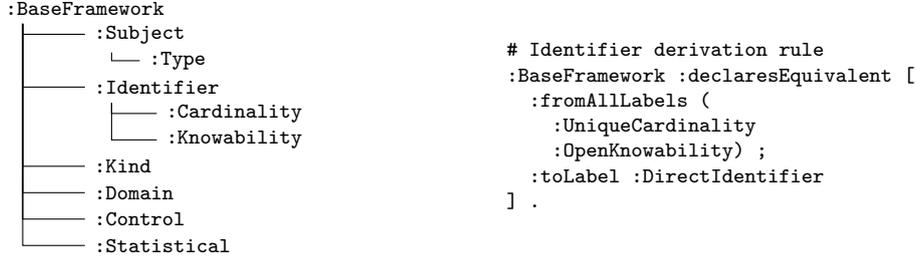
\begin{figure}[ht]
\centering
\begin{minipage}[c]{0.48\textwidth}
\raggedright
\begin{tikzpicture}[dirtree]
\node {:BaseFramework}
  child { node {:Subject}
    child { node {:Type} }
  }
  child { node {:Identifier}
    child { node {:Cardinality} }
    child { node {:Knowability} }
  }
  child { node {:Kind} }
  child { node {:Domain} }
  child { node {:Control} }
  child { node {:Statistical} };
\end{tikzpicture}
\end{minipage}%
\hspace{0.5em}%
\begin{minipage}[c]{0.50\textwidth}
\begin{lstlisting}[basicstyle=\ttfamily\scriptsize, frame=none]
# Identifier derivation rule
:BaseFramework :declaresEquivalent [
  :fromAllLabels (
    :UniqueCardinality
    :OpenKnowability) ;
  :toLabel :DirectIdentifier
] .
\end{lstlisting}
\end{minipage}
\caption{Base framework facet taxonomy (left) and an identifier derivation rule (right).}
\label{fig:base-framework-structure}
\end{figure}

\paragraph{Identifier Derivation.}
The \facet{Identifier} facet derives types by combining \clabel{Cardinality} (\clabel{Unique}, \clabel{ClosedGroup}, \clabel{OpenGroup}) with \clabel{Knowability} (\clabel{Open}, \clabel{Closed}), as shown in Table~\ref{tab:identifier-derivation}. This yields nuanced classifications: SSNs, with \clabel{UniqueCardinality} and \clabel{OpenKnowability}, become \clabel{DirectIdentifier}, while internal UUIDs with \clabel{ClosedKnowability} are classified as \clabel{InternalIdentifier}.

\begin{table}[ht]
\centering
\small
\setlength{\tabcolsep}{4pt}
\renewcommand{\arraystretch}{0.75}
\begin{tabular}{lclcl}
\toprule
\textbf{Derived Label} & & \textbf{Cardinality} & & \textbf{Knowability} \\
\midrule
\clabel{DirectIdentifier} & $\equiv$ & \clabel{Unique} & $\land$ & \clabel{Open} \\
\clabel{InternalIdentifier} & $\equiv$ & \clabel{Unique} & $\land$ & \clabel{Closed} \\
\clabel{IndirectIdentifier} & $\equiv$ & \clabel{OpenGroup} & $\land$ & \clabel{Open} \\
\bottomrule
\end{tabular}
\caption{Identifier classifications derived from orthogonal properties.}
\label{tab:identifier-derivation}
\end{table}
\vspace{-\baselineskip}

\paragraph{HIPAA.}
This framework extends the Base \facet{Control} facet to implement the Privacy Rule, introducing concepts centered on Protected Health Information (PHI). It applies expansive propagation: \emph{inward} (to contained fields), \emph{outward} (containers containing PHI are themselves classified as PHI), \emph{peer} (to sibling containers), and \emph{joinable} (to related data). Healthcare data becomes PHI through conditional implication: any container labeled \clabel{Healthcare} that contains either \clabel{HIPAAIdentifier} or \clabel{ProtectedHealthInformation} receives PHI status.

\paragraph{De-identification.}
Two sub-frameworks define de-identification pathways. \emph{Safe Harbor} removes $18$ enumerated identifiers mapped to \clabel{SafeHarborIdentifier} in healthcare domains. \emph{Expert Determination} requires statistical evidence of ``very small'' re-identification risk, leveraging Base's \clabel{KAnonymityAnalysis} assertions with \texttt{minimumCohortSize} parameters. When $k < 3$ (configurable), \clabel{HighReidentificationRisk} triggers PHI classification, enabling tunable risk thresholds. Figure~\ref{fig:hipaa-deid} enumerates all 18 identifier types and shows Expert Determination's threshold comparison structure.

\begin{figure}[ht]
\centering
\begin{minipage}[t]{0.42\textwidth}
\begin{lstlisting}[basicstyle=\ttfamily\scriptsize, frame=none]
:HIPAASafeHarborFramework
  :declaresSubclassOf [
    :fromAnyLabel
      b:Name,
      b:Address,
      b:MomentData,
      b:Phone,
      b:Fax,
      b:Email,
      b:SSN,
      b:MedicalRecordNumber,
      b:HealthPlanNumber,
      b:AccountNumber,
      b:CertificateNumber,
      b:VehicleIdentifier,
      b:DeviceIdentifier,
      b:WebURL,
      b:IPAddress,
      b:BiometricData,
      b:FaceImage,
      b:UniqueID ;
    :toLabel
      :SafeHarborIdentifier
  ] .
\end{lstlisting}
\end{minipage}%
\hspace{0.3em}\textcolor{gray!50}{\vrule}\hspace{0.3em}%
\begin{minipage}[t]{0.56\textwidth}
\begin{lstlisting}[basicstyle=\ttfamily\scriptsize, frame=none, xleftmargin=0.25cm]
:HIPAAExpertDeterminationFramework
  :declaresImplication [
    a :ConditionalImplication ;
    :fromLabel b:KAnonymityAnalysis ;
    :toLabel :HighReidentificationRisk ;
    :hasCondition [
      a :ComparisonCondition ;
      :leftSource [
        :sourceType :LabelParameter ;
        :sourceLabel b:KAnonymityAnalysis ;
        :sourceParameter "minimumCohortSize"
      ] ;
      :rightSource [
        :sourceType :LabelParameter ;
        :sourceLabel :ExpertDeterminationThreshold ;
        :sourceParameter "kThreshold" ;
        :defaultValue 3
      ] ;
      :comparisonOperator :lessThan
    ]
  ] .
\end{lstlisting}
\end{minipage}
\caption{HIPAA de-identification frameworks.\footnotemark\ Safe Harbor (left) maps 18 identifier types to SafeHarborIdentifier. Expert Determination (right) flags high re-identification risk when $k < 3$.}
\label{fig:hipaa-deid}
\end{figure}
\footnotetext{Throughout, code listings abbreviate the \texttt{pj:} (Parajudica) namespace as \texttt{:} and \texttt{base:} as \texttt{b:}.}

\paragraph{GDPR.}
This framework distinguishes data \emph{about} individuals from data that \emph{identifies} them~\cite{wp29-personaldata}. The \clabel{Individual} label marks persistent characteristics, while \clabel{PersonalData} requires linkage with identifiers. Removing identifiers preserves \clabel{Individual} but eliminates \clabel{PersonalData}. GDPR's hierarchy nests health data within special category data, itself within personal data. Medical codes become health data only with identifiers in healthcare contexts, keeping anonymous research data unclassified. Propagation is strictly \emph{inward}, maintaining field-level precision versus HIPAA's eager propagation. Figure~\ref{fig:gdpr-ema} shows GDPR's encoding alongside EMA, which extends it with a stricter $k < 12$ threshold for public release.

\begin{figure}[ht]
\centering
\begin{minipage}[t]{0.47\textwidth}
\begin{lstlisting}[basicstyle=\ttfamily\scriptsize, frame=none]
:GDPRFramework
:declaresImplication [
  a :ConditionalImplication ;
  :fromLabel b:Individual ;
  :toLabel :PersonalData ;
  :hasCondition [
    a :ContainsLabelCondition ;
    :requiresContains b:IdentifierData
  ]
] .

:GDPRFramework
:declaresSubclassOf [
  :fromLabel :DataConcerningHealth ;
  :isSubclassOf :SpecialCategoryData
] .

:GDPRFramework
:declaresPropagation [
  :propagatesLabel :PersonalData ;
  :propagationDirection :Inward
] .
\end{lstlisting}
\end{minipage}%
\hspace{0.3em}\textcolor{gray!50}{\vrule}\hspace{0.3em}%
\begin{minipage}[t]{0.50\textwidth}
\begin{lstlisting}[basicstyle=\ttfamily\scriptsize, frame=none, xleftmargin=0.25cm]
:EMAFramework a :Framework ;
  :extends gdpr:GDPRFramework .

:EMAFramework
:declaresImplication [
  a :ConditionalImplication ;
  :fromLabel b:KAnonymityAnalysis ;
  :toLabel :HighReidentificationRisk ;
  :hasCondition [
    a :ComparisonCondition ;
    :leftSource [
      :sourceLabel b:KAnonymityAnalysis ;
      :sourceParameter "minimumCohortSize"
    ] ;
    :rightSource [
      :defaultValue 12
    ] ;
    :comparisonOperator :lessThan
  ]
] .
\end{lstlisting}
\end{minipage}
\caption{GDPR (left) requires identifiers for personal data classification, establishes health data hierarchy, and propagates inward. EMA (right) extends GDPR with $k < 12$ threshold for high re-identification risk.}
\label{fig:gdpr-ema}
\end{figure}

\paragraph{Italian DPA.}
The Italian DPA applies a stricter GDPR interpretation, treating \emph{any} unique information about individuals as personal data. While GDPR requires both cardinality and knowability for identifiability, the Italian DPA considers uniqueness alone sufficient (Figure~\ref{fig:italian-dpa}). Thus internal UUIDs with \clabel{ClosedKnowability} become \clabel{PersonalData} under Italian rules.

\section{Validation}\label{sec:validation}

We validate our metamodel against the five compliance challenges from Section~\ref{sec:legal} through a healthcare scenario where a clinic maintains employee vaccination records separately from patient treatment data, then joins them for research purposes.

\begin{figure}[ht]
\centering
\begin{tikzpicture}[
    table/.style={draw, fill=white, rounded corners=2pt, font=\scriptsize,
                  minimum width=1.8cm, minimum height=0.7cm, align=center},
    scope region/.style={draw, rounded corners},
    scope label/.style={font=\scriptsize},
    jline/.style={->, gray}
]
\node[table] (PI) at (0,0) {ProvidersInfo};

\node[table] (PtInfo) at (5,1) {PatientInfo};
\node[table] (PtEnc) at (5,0) {PatientEncounters};
\node[table] (PtTreat) at (5,-1) {PatientTreatments};

\draw[jline] (PI.east) -- (PtEnc.west);
\draw[jline] (PI.east) -- (PtTreat.west);

\draw[jline] (PtInfo.south) -- (PtEnc.north);
\draw[jline] (PtInfo.east) -- ++(0.5,0) |- (PtTreat.east);
\draw[jline] (PtEnc.south) -- (PtTreat.north);

\begin{scope}[on background layer]
  \node[scope region, dashed, fit=(PI)(PtInfo)(PtTreat), inner sep=18pt] (RS) {};
  \node[scope region, fit=(PI), inner sep=12pt, fill=gray!10] (HR) {};
  \node[scope region, fit=(PtInfo)(PtEnc)(PtTreat), inner sep=12pt, fill=gray!10] (MS) {};
\end{scope}
\node[scope label, anchor=north west] at (RS.north west) {ResearchScope};
\node[scope label, anchor=south west] at (HR.south west) {HRScope};
\node[scope label, anchor=south east] at (MS.south east) {MedicalScope};

\end{tikzpicture}
\caption{Healthcare scenario. Tables exist once with ground assertions (intrinsic properties); scopes define visibility for HR, clinical, and research contexts respectively. Compliance labels are derived independently per scope based on visible tables: HRScope sees only ProvidersInfo; MedicalScope sees only the patient tables; ResearchScope (dashed) sees both, enabling cross-system joins that derive different labels than in isolated scopes. Arrows indicate \texttt{:joinableWith} declarations.}
\label{fig:scenario}
\end{figure}
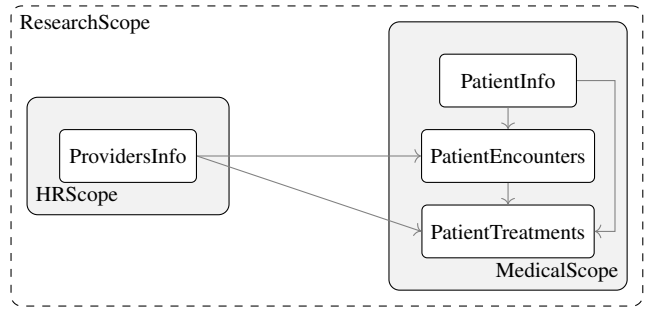

\paragraph{Scenario Setup.}
A clinic's employee vaccination records (\clabel{ProvidersInfo}) exist in \sym{HRScope} containing names, SSNs, and vaccination dates. Patient treatment data (\clabel{PatientTreatments}) resides in \sym{MedicalScope} with diagnoses and procedures. For research analysis, both datasets become available in \sym{ResearchScope} where they can be joined. The same \clabel{ProvidersInfo} data thus appears in multiple scopes with different available relationships. Figure~\ref{fig:scenario} illustrates this scenario.

\paragraph{Challenge 1: Context-dependent classification.}
Employee vaccination records demonstrate how identical data receives different statuses across environments. In \sym{HRScope}, \clabel{ProvidersInfo} lacks healthcare context despite containing names and SSNs: HIPAA's conditional rule requires both identifiers and healthcare context for PHI classification. In \sym{ResearchScope}, the same records acquire PHI status through joinable propagation from \clabel{PatientTreatments}, mirroring how vaccination records are treated as HR data in one setting but PHI in clinical systems.

\paragraph{Challenge 2: Diverging interpretations.}
When joined, employee-patient records produce contradictory outcomes across frameworks. HIPAA treats the entire joined dataset as PHI through expansive propagation: employee names and SSNs acquire PHI status via join, vaccination dates become PHI due to healthcare context, resulting in the entire join product being classified as PHI. In contrast, GDPR maintains field-level distinctions: employee names and SSNs remain PersonalData, vaccination dates are classified as HealthData, and the join product preserves these mixed types rather than collapsing to a single classification. This demonstrates how frameworks split the same data differently.

\paragraph{Challenge 3: Proximity semantics.}
The availability of nearby data within a scope alters compliance status. In \sym{ResearchScope}, employee identifiers acquire PHI status when linked to patient treatments through joinable relationships. HIPAA's four propagation patterns (\emph{inward}, \emph{outward}, \emph{peer}, \emph{joinable}) ensure comprehensive PHI spread, while GDPR's \emph{inward}-only propagation preserves field-level precision: \clabel{PatientTreatments} carries \clabel{HealthData} while sibling \clabel{ProvidersInfo} remains \clabel{PersonalData}.

\paragraph{Challenge 4: De-identification scenarios.}
The metamodel accommodates both structural and statistical de-identification approaches. HIPAA Safe Harbor requires removing enumerated identifiers (the 18 specific types). Expert Determination uses statistical assertions: when $k < 3$, \clabel{HighReidentificationRisk} triggers PHI classification. EMA requires $k \geq 12$ for public release. The Italian DPA rejects risk-based methods altogether, treating even internal UUIDs with \clabel{ClosedKnowability} as \clabel{PersonalData} based solely on uniqueness.

\paragraph{Challenge 5: Comparative evaluation.}
Side-by-side analysis makes divergences explicit through framework attribution via \sym{assertedByFramework}. \clabel{ProvidersInfo} simultaneously carries PHI status (HIPAA in \sym{ResearchScope}) and PersonalData (GDPR in both scopes). Like Table~\ref{tab:regulatory-divergence} showing a $k=5$ dataset acceptable under one framework but prohibited under others, our validation demonstrates how the metamodel enables systematic comparison of framework determinations.

\section{Formalization}\label{sec:formalization}

We formalize our metamodel using first-order logic with least fixed-point operator (FO+LFP), establishing polynomial-time complexity bounds for compliance inference.

\paragraph{Compliance Metamodel.}
A \emph{compliance metamodel} is a tuple $(X,L,F,\prec)$, consisting of facets $X$, labels $L$, frameworks $F$, and framework inheritance $\prec$ (an acyclic relation on frameworks with edges from parent to child). A \emph{compliance environment} is a tuple $(D,G,A_0,\sqsubset,\bowtie)$, consisting of containers $D$, scopes $G$, initial assertions $A_0$, immediate containment $\sqsubset$ (irreflexive), and joinability $\bowtie$ (a symmetric relation on containers).

Our goal is to compute the complete assertion set $A\subseteq U:=D\times L\times G\times F$ given a compliance metamodel and an environment. An assertion $(d,l,g,f)$ assigns label $l$ to container $d$ in scope $g$ under framework $f$. Initially $A_0\subseteq A$; inference derives all consequences to obtain $A$.

\paragraph{Notation.} Write $l\in_f[d]_g$ for $(d,l,g,f)\in A$. From $\sqsubset$ we derive the following canonical relations: $\textsf{child}(x,y) := (x\sqsubset y)$, $\textsf{parent}(x,y) := (y\sqsubset x)$, $\textsf{desc}(x,y) := (x\sqsubset^+ y)$, $\textsf{sib}(x,y) := (x\!\ne\! y\ \wedge\ \exists p\,(\textsf{child}(x,p)\wedge \textsf{child}(y,p)))$, and $\textsf{id}(x,y) := (x=y)$. The joinability relation gives $\textsf{joinable}(x,y) := (x \bowtie y)$. The relation $\sqsubset^+$ denotes the transitive closure of $\sqsubset$, capturing indirect containment where $x$ is a descendant of $y$ through one or more intermediate containers.

\paragraph{Framework inheritance.}
Each framework $f\in F$ declares rules $R_f^{\mathrm{decl}}$. A child $f$ \emph{overrides} inherited rules with head $l$ iff it declares any rule with head $l$. With $\prec$ acyclic, the effective rules for $f$ are
\[
R_f = R_f^{\mathrm{decl}}\ \cup\ \bigcup_{f'\prec f}\{r\in R_{f'}:\neg\,\mathrm{overrides}(f,r)\},\qquad
R=\bigcup_{f\in F}R_f.
\]

\paragraph{Inference rules.}
For a framework $f \in F$, $R_f$ partitions into \begin{enumerate*}[label=(\roman*)] \item simple rules $R_f^s\subseteq L\times L$ of the form $l_1\to l_2$, \item conditional rules $R_f^c\subseteq L\times\Gamma\times L$ of the form $l_1\xrightarrow{\phi} l_2$, \item pure implication rules $R_f^p\subseteq\Gamma\times L$ of the form $\phi\Rightarrow l$, and \item propagation rules $R_f^t\subseteq\{\textsf{child},\textsf{parent},\textsf{sib},\textsf{joinable}\}\times L$ of the form $(\sigma,l)$.\end{enumerate*} One-step derivability $A\vdash\tau$ is given by:
\[
\footnotesize
\frac{l_1\in[d]_g,\ (l_1\to l_2)\in R_f^s}{A\vdash(d,l_2,g,f)}\quad
\frac{l_1\in[d]_g,\ (l_1\xrightarrow{\phi} l_2)\in R_f^c,\ \phi(d,g)}{A\vdash(d,l_2,g,f)}\quad
\frac{(\phi\Rightarrow l)\in R_f^p,\ \phi(d,g)}{A\vdash(d,l,g,f)}
\]
\[
\small
\frac{l\in_f[d_1]_g,\ \sigma(d_1,d_2),\ (\sigma,l)\in R_f^t}{A\vdash(d_2,l,g,f)}.
\]
The condition language $\Gamma$ is closed under $\land,\lor$, generated by atoms $\textsf{HasLabel}_{\sigma,l}(d,g)$ (defined as $\exists d'\,(\sigma(d',d)\wedge l\in[d']_g)$ for $l\in L$ and $\sigma\in\{\textsf{id},\textsf{child},\textsf{parent},\textsf{desc},\textsf{sib}\}$) and what amounts to extensional predicates $\textsf{P}(d,g)\in\mathcal{P}$. Here, $\mathcal{P}$ is a fixed finite set of base predicates from the environment.

\begin{theorem}[Fixed point and complexity]\label{thm:convergence}
Define the immediate-consequence operator $T(A)=A\cup\{\tau: A\vdash\tau\}$. Starting from $A^{(0)}=A_0$, iterate $A^{(i+1)}=T(A^{(i)})$. Since $U=D\times L\times G\times F$ is finite and all rules are positive, the sequence reaches the least fixed point $A^*$ in at most $|U|$ iterations. Moreover, it converges in polynomial time (naively, each round considers $O(|R|\,\Delta\,|U|)$ groundings, where $\Delta$ bounds the out-degree of the propagation relations).
\end{theorem}

Table~\ref{tab:rdf-mapping} shows how the formal structures map to their RDF implementation.

\begin{table}[ht]
\centering
\small
\setlength{\tabcolsep}{3pt}
\renewcommand{\arraystretch}{0.85}
\begin{tabular}{@{}rl@{\hspace{0.5cm}}rl@{}}
\toprule
\multicolumn{2}{c}{\textbf{Primary Structures}} & \multicolumn{2}{c}{\textbf{Rule-Supporting Structures}} \\
\midrule
$D$ & {\footnotesize\texttt{:DataContainer}} & $\text{hasLabel}(d,l,g)$ & \fnsym{HasLabelCondition} \\
$L$ & {\footnotesize\texttt{:ComplianceLabel}} & $\text{containsLabel}(d,l,g)$ & \fnsym{ContainsLabelCondition} \\
$F$ & {\footnotesize\texttt{:Framework}} & container relations & \fnsym{RelationLabelCondition} \\
$G$ & {\footnotesize\texttt{:GovernanceScope}} & $\text{param}(d,\sigma,\text{op},\theta)$ & \fnsym{ParameterCheckCondition} \\
$X$ & {\footnotesize\texttt{:Facet}} & $\phi_1 \land \phi_2, \phi_1 \lor \phi_2$ & \fnsym{CompositeCondition} \\
\cmidrule{3-4}
$d_1 \sqsubset d_2$ & {\footnotesize\texttt{d1 :contains d2}} & Rule declarations: & \fnsym{SubclassDeclaration} \\
$d_1 \bowtie d_2$ & {\footnotesize\texttt{d1 :joinableWith d2}} & & \fnsym{ConditionalEquivalence} \\
$(d,l,g,f) \in A$ & RDF assertions (Figure~\ref{fig:rdf-structures}) & & \fnsym{PropagationDeclaration} \\
\bottomrule
\end{tabular}
\caption{Mapping from formal model to RDF implementation.}
\label{tab:rdf-mapping}
\end{table}

\begin{figure}[ht]
\centering
\small
\begin{tabular}{@{}rl@{\hspace{1.2em}}rl@{\hspace{1.2em}}rl@{}}
\toprule
\multicolumn{2}{c}{\textbf{ComplianceAssertion}} & \multicolumn{2}{c}{\textbf{ContainmentAssertion}} & \multicolumn{2}{c}{\textbf{ConditionEvaluation}} \\
\midrule
\texttt{assertedOn} & $D$ & \texttt{assertedOn} & $D$ & \texttt{evaluatesCondition} & $\Gamma$ \\
\texttt{assertsLabel} & $L$ & \texttt{assertsLabel} & $L$ & \texttt{evaluatedOn} & $D$ \\
\texttt{assertedInScope} & $G$ & \texttt{assertedInScope} & $G$ & \texttt{evaluatedInScope} & $G$ \\
\texttt{byFramework} & $F$ & & & \texttt{evaluationResult} & $\mathbb{B}$ \\
\texttt{hasParameter}* & & & & & \\
\bottomrule
\end{tabular}
\caption{RDF properties for assertion types, with ranges from the formal model. Asterisk (*) denotes optional.}
\label{fig:rdf-structures}
\end{figure}

\section{Implementation}\label{sec:implementation}

The metamodel is implemented as a Python package using Oxigraph~\cite{oxigraph} as the underlying RDF triple store. The complete implementation and example frameworks are available as open source~\cite{parajudica}. The package provides a CLI entry point ({\footnotesize\texttt{cli.py}}) and a programmatic API via the {\footnotesize\texttt{InferenceSystem}} class ({\footnotesize\texttt{engine.py}}). Supporting modules handle Jena rule compilation ({\footnotesize\texttt{jena\_compiler.py}}), triple store operations ({\footnotesize\texttt{oxigraph\_runner.py}}), blank node skolemization ({\footnotesize\texttt{skolemizer.py}}), and result caching ({\footnotesize\texttt{cache.py}}).

Frameworks are defined by manifest files ({\footnotesize\texttt{framework.toml}}) declaring metadata, dependencies, type ({\footnotesize\texttt{internal}}, {\footnotesize\texttt{core}}, {\footnotesize\texttt{privacy}}, or {\footnotesize\texttt{custom}}), and three categories of inputs: \emph{models} (TTL files defining vocabularies, labels, and facets), \emph{rules} (Jena files compiled to SPARQL at startup), and \emph{constructs} (SPARQL CONSTRUCT queries executed during iteration). Users supply \emph{environment} data (TTL files describing the data landscape) and \emph{queries} (SPARQL SELECT for extracting results); an optional \emph{cache} can bypass loading with a pre-computed store.

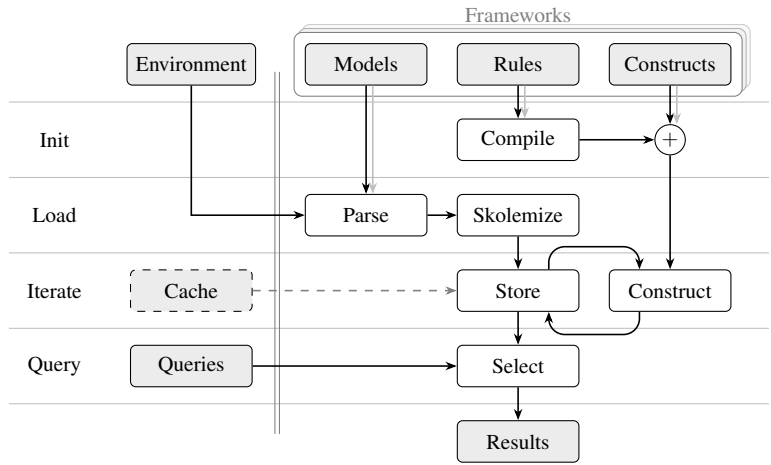
\begin{figure}[ht]
\centering
\begin{tikzpicture}[
    box/.style={draw, rounded corners=2pt, minimum width=1.6cm, minimum height=0.55cm, font=\scriptsize, align=center},
    input/.style={box, fill=gray!15},
    stage/.style={box, fill=white},
    arr/.style={-{Stealth[length=1.5mm]}, semithick},
    cachearr/.style={-{Stealth[length=1.5mm]}, semithick, dashed, gray},
    phaseline/.style={draw, gray!50},
    phaselabel/.style={font=\scriptsize},
    junction/.style={draw, circle, minimum size=0.4cm, inner sep=0pt, font=\scriptsize},
]
    \node[input] (models) at (1.5, -0.3) {Models};
    \node[input] (rules) at (3.5, -0.3) {Rules};
    \node[input] (constructs) at (5.5, -0.3) {Constructs};
    \begin{scope}[on background layer]
        \node[draw, rounded corners=3pt, gray!40, fill=gray!5, fit=(models)(rules)(constructs), inner sep=4pt, xshift=3pt, yshift=3pt] {};
        \node[draw, rounded corners=3pt, gray!60, fill=gray!10, fit=(models)(rules)(constructs), inner sep=4pt, xshift=1.5pt, yshift=1.5pt] {};
        \node[draw, rounded corners=3pt, gray, fill=white, fit=(models)(rules)(constructs), inner sep=4pt, label={[font=\scriptsize, text=gray]above:Frameworks}] {};
    \end{scope}

    \node[input] (env) at (-0.8, -0.3) {Environment};

    \draw[phaseline] (-3.2, -0.8) -- (6.8, -0.8);
    \node[phaselabel] at (-2.6, -1.3) {Init};
    \node[stage] (compile) at (3.5, -1.3) {Compile};
    \node[junction] (sum) at (5.5, -1.3) {$+$};

    \draw[phaseline] (-3.2, -1.8) -- (6.8, -1.8);
    \node[phaselabel] at (-2.6, -2.3) {Load};
    \node[stage] (load) at (1.5, -2.3) {Parse};
    \node[stage] (skolem) at (3.5, -2.3) {Skolemize};

    \draw[phaseline] (-3.2, -2.8) -- (6.8, -2.8);
    \node[phaselabel] at (-2.6, -3.3) {Iterate};
    \node[input, dashed] (cache) at (-0.8, -3.3) {Cache};
    \node[stage] (store) at (3.5, -3.3) {Store};
    \node[stage] (construct) at (5.5, -3.3) {Construct};

    \draw[phaseline] (-3.2, -3.8) -- (6.8, -3.8);
    \node[phaselabel] at (-2.6, -4.3) {Query};
    \node[stage] (select) at (3.5, -4.3) {Select};
    \node[input] (qry) at (-0.8, -4.3) {Queries};

    \draw[phaseline] (-3.2, -4.8) -- (6.8, -4.8);
    \node[input] (results) at (3.5, -5.3) {Results};

    \draw[gray] (0.30, -0.4) -- (0.30, -5.2);
    \draw[gray] (0.36, -0.4) -- (0.36, -5.2);

    \draw[arr] (models) -- (load);
    \draw[arr, gray!50, transform canvas={xshift=2.5pt}] (models) -- (load);
    \draw[arr] (rules) -- (compile);
    \draw[arr, gray!50, transform canvas={xshift=2.5pt}] (rules) -- (compile);
    \draw[arr] (compile) -- (sum);
    \draw[arr] (constructs) -- (sum);
    \draw[arr, gray!50, transform canvas={xshift=2.5pt}] (constructs) -- (sum);
    \draw[arr] (sum) -- (construct);

    \draw[arr] (env) |- (load.west);

    \draw[arr] (load) -- (skolem);
    \draw[arr] (skolem) -- (store);
    \draw[arr] (store) -- (select);
    \draw[arr] (select) -- (results);

    \draw[arr, rounded corners=6pt] ($(store.north east)!0.5!(store.north)$) -- ++(0, 0.3) -| ($(construct.north west)!0.5!(construct.north)$);
    \draw[arr, rounded corners=6pt] ($(construct.south west)!0.5!(construct.south)$) -- ++(0, -0.3) -| ($(store.south east)!0.5!(store.south)$);

    \draw[arr] (qry.east) -- (select.west);

    \draw[cachearr] (cache) -- (store);
\end{tikzpicture}
\caption{Inference pipeline. User inputs (left) include environment data, an optional cache, and queries. Frameworks (right, stacked to show multiplicity) contribute models, rules, and constructs. The four phases are: \emph{Init} compiles Jena rules and merges them with declared constructs; \emph{Load} parses TTL files and skolemizes blank nodes; \emph{Iterate} executes CONSTRUCT queries in a fixed-point loop; \emph{Query} runs SELECT queries to produce results. A cached store can bypass loading.}
\label{fig:architecture}
\end{figure}

The system operates in four phases: \emph{init}, \emph{load}, \emph{iterate}, and \emph{query}. Figure~\ref{fig:architecture} illustrates the pipeline.

\paragraph{Init.}
At initialization, the engine discovers frameworks via manifest files, loading them in dependency order. Jena rules~\cite{jena} are compiled into SPARQL CONSTRUCT queries (Figure~\ref{fig:jena-compile}): a rule {\footnotesize\texttt{[Name: (body) -> (head)]}} becomes a query matching body patterns and constructing head triples. Built-ins translate to SPARQL equivalents: {\footnotesize\texttt{noValue}} becomes {\footnotesize\texttt{FILTER NOT EXISTS}}, {\footnotesize\texttt{makeSkolem}} becomes {\footnotesize\texttt{BIND(IRI(CONCAT(...)))}}. Compilation occurs once at startup; the compiled queries join framework-declared SPARQL constructs for execution.

\begin{figure}[ht]
\centering
\begin{minipage}[t]{0.50\textwidth}
\begin{lstlisting}[basicstyle=\ttfamily\scriptsize, frame=none]
[InheritPropagationRules:
  (?C :extends ?P)
  (?P :hasPropagationRule ?R)
  noValue(?C :hasPropagationRule ?R)
  ->
  (?C :hasPropagationRule ?R)
]
\end{lstlisting}
\end{minipage}%
\begin{minipage}[t]{0.50\textwidth}
\begin{lstlisting}[basicstyle=\ttfamily\scriptsize, frame=none, language=SPARQL]
CONSTRUCT
  { ?C :hasPropagationRule ?R }
WHERE {
  ?C :extends ?P .
  ?P :hasPropagationRule ?R .
  FILTER NOT EXISTS
    { ?C :hasPropagationRule ?R }
}
\end{lstlisting}
\end{minipage}
\caption{Jena rule (left) compiled to SPARQL CONSTRUCT (right). Child framework $C$ inherits propagation rule $R$ from parent $P$ unless already defined; \texttt{noValue} becomes \texttt{FILTER NOT EXISTS}.}
\label{fig:jena-compile}
\end{figure}

\paragraph{Load.}
The engine loads TTL files in framework dependency order (internal metamodel, core frameworks, privacy frameworks, user data). Each file is parsed into the Oxigraph store with a distinct base IRI. Initially, blank nodes from different files have independent scopes: a {\footnotesize\texttt{\_:b0}} in one file is unrelated to {\footnotesize\texttt{\_:b0}} in another. Before iteration begins, all blank nodes are replaced with stable, content-hashed URIs. The skolemizer computes a signature from each blank node's properties (predicates and objects), then generates a deterministic URI via SHA-256 hashing. This unification across serialization boundaries is essential: framework declarations use inline blank nodes (e.g., {\footnotesize\texttt{:Framework :declaresSubclassOf [ :fromLabel :X ; :toLabel :Y ]}}), and inheritance rules must match parent declarations loaded from separate files. Content-based skolemization ensures structurally identical declarations receive the same URI regardless of source, enabling correct rule inheritance.

\paragraph{Iterate.}
The engine enters a fixed-point loop, executing all SPARQL CONSTRUCT queries each round. When a query's WHERE clause matches patterns in the store, its CONSTRUCT clause generates new triples that are added to the store. The loop terminates when a complete round produces no new triples, guaranteeing convergence per Theorem~\ref{thm:convergence}. Each round processes frameworks in dependency order, executing all construct queries from each framework's manifest. Frameworks may declare additional domain-specific constructs beyond those provided by the core metamodel.

The core metamodel ({\footnotesize\texttt{pj}}) provides constructs implementing the formal semantics from Section~\ref{sec:formalization}. These operations execute every round; the fixed-point loop allows us to proceed incrementally which can loosely be organized into the following functions:
\begin{enumerate}[leftmargin=2em,itemsep=2pt]
\item \textbf{Indexing} builds lookup structures (e.g., {\footnotesize\texttt{:hasLabelInScope}}) for efficient pattern matching.

\item \textbf{Expansion} transforms compact declarations into executable form: {\footnotesize\texttt{fromAnyLabel}} lists expand to individual rules, facet-level propagation rules expand to labels, equivalences generate bidirectional implications.

\item \textbf{Materialization} pre-computes derived structures; containment assertions materialize ``container $C$ contains label $L$'' for efficient condition checking.

\item \textbf{Marking} identifies conditions requiring evaluation (when a triggering label exists but its target is absent), propagating marks through composite conditions.

\item \textbf{Evaluation} checks marked conditions: atomic conditions ({\footnotesize\texttt{Contains}}, {\footnotesize\texttt{Relation}}, {\footnotesize\texttt{Comparison}}) first, then composites (AND/OR), recording outcomes as {\footnotesize\texttt{:ConditionEvaluation}} triples.

\item \textbf{Derivation} fires implications and propagation rules based on condition outcomes, spreading labels \emph{inward}, \emph{outward}, among \emph{peers}, or across \emph{joinable} relationships according to framework-specific patterns.
\end{enumerate}
Each derived assertion receives a deterministic URI via {\footnotesize\texttt{BIND(IRI(CONCAT(...)))}} templates encoding container, label, framework, and scope, ensuring idempotent operations across iterations. Figure~\ref{fig:and-condition} shows a composite AND condition query.

\begin{figure}[ht]
\centering
\begin{minipage}[t]{0.46\textwidth}
\vspace{0pt}
\begin{lstlisting}[basicstyle=\ttfamily\scriptsize, frame=none]
:ItalianDPAFramework a :Framework ;
  :extends gdpr:GDPRFramework .

:ItalianDPAFramework
:declaresImplication [
  a :ConditionalImplication ;
  :fromLabel b:Individual ;
  :toLabel gdpr:PersonalData ;
  :hasCondition [
    a :ContainsLabelCondition ;
    :requiresContains b:UniqueCardinality
  ]
] .

:ItalianDPAFramework
:declaresImplication [
  a :ConditionalImplication ;
  :fromLabel b:UniqueCardinality ;
  :toLabel gdpr:PersonalData ;
  :hasCondition [
    a :RelationLabelCondition ;
    :onRelation :Self ;
    :requiresLabel b:Individual
  ]
] .
\end{lstlisting}
\end{minipage}%
\hspace{0.3em}\textcolor{gray!50}{\vrule}\hspace{0.3em}%
\begin{minipage}[t]{0.50\textwidth}
\vspace{0pt}
\begin{lstlisting}[basicstyle=\ttfamily\scriptsize, frame=none, xleftmargin=0.25cm]
CONSTRUCT {
  ?count a :ConditionCountAssertion ;
    :forCondition ?andCond ;
    :onContainer ?c ; :inScope ?g ;
    :satisfiedCount ?satisfied ;
    :totalCount ?total .
}
WHERE {
  SELECT ?andCond ?c ?g
         (COUNT(?satEval) as ?satisfied)
         (COUNT(?subCond) as ?total)
  WHERE {
    ?andCond a :CompositeCondition ;
             :logicalOperator :AND ;
             :hasCondition ?subCond .
    ?eval :evaluatesCondition ?andCond ;
          :evaluatedOn ?c ;
          :evaluatedInScope ?g .
    OPTIONAL {
      ?satEval :evaluatesCondition ?subCond ;
               :evaluatedOn ?c ;
               :evaluationResult true .
    }
  }
  GROUP BY ?andCond ?c ?g
}
\end{lstlisting}
\end{minipage}
\caption{(a)~Italian DPA Framework: uniqueness triggers personal data classification.\label{fig:italian-dpa} (b)~AND condition evaluation via subcondition counting.\label{fig:and-condition}}
\label{fig:code-examples}
\end{figure}

\paragraph{Optimization.}
Several techniques ensure scalability. Deterministic URI generation prevents duplicate derivations that would arise from SPARQL's fresh blank node semantics. The marking phase avoids evaluating conditions whose outcomes are irrelevant. Pre-computed containment assertions eliminate recursive traversal during condition checking. Separate query paths for conditional and unconditional rules skip condition evaluation when unnecessary. Statistical analyses like $k$-anonymity use SPARQL UPDATE queries to compute minimum cohort sizes $k = \min_{g \in G} |g|$ where $G$ partitions records by quasi-identifiers.

\section{Related Work and Conclusion}\label{sec:related-conclusion}

The Data Privacy Vocabulary (DPV)~\cite{pandit2019dpv} and the Open Digital Rights Language (ODRL)~\cite{odrl-model-2018,odrl-vocab-2018} are the leading semantic standards for privacy and governance. DPV, developed by the Data Privacy Vocabularies and Controls Community Group, provides a cross-jurisdictional vocabulary for describing data processing and is incorporated into ISO/IEC TS 27560:2023~\cite{iso27560}, which standardizes consent records using concepts such as \texttt{dpv:hasPurpose} and \texttt{dpv:ServiceProvision}. ODRL specifies permissions, prohibitions, and duties for data usage and underpins initiatives like the International Data Spaces Association~\cite{bader2020}. Both vocabularies are descriptive: DPV documents compliance attributes (e.g., \texttt{dpv:SensitivePersonalData}), while ODRL defines policies referencing them. ODRL supports temporal, spatial, and purpose-based constraints but assumes compliance labels are static. Other related efforts include LegalRuleML~\cite{legalruleml} (legal norms), PROV-O~\cite{prov-o} (provenance of compliance labels), and DCAT~\cite{dcat} (dataset description, aligned with our notion of data containers but lacking propagation). In domains, FIBO~\cite{fibo} and Gist~\cite{gist} provide complementary vocabularies, while International Data Spaces demonstrates ODRL integration with domain-specific vocabularies, offering a precedent for layering computational semantics onto existing standards. 

Against this backdrop, we introduced Parajudica, a metamodel for multi-framework compliance addressing the five challenges in Section~\ref{sec:legal}. Governance scopes capture context-dependent classification, propagation rules reflect divergent philosophies (HIPAA’s expansive vs.\ GDPR’s precise), parameterized assertions encode quantitative measures such as $k$-anonymity, and framework attribution preserves provenance. The formalization establishes polynomial-time bounds via stratified fixed-point computation. Validation in a healthcare scenario showed vaccination records acquire PHI status only when joined with patient data, and an RDF/SPARQL implementation demonstrates feasibility atop existing semantic technologies. 

Future work includes temporal reasoning for evolving states (including condition removal), PROV-O integration for audit trails, and comparative analysis to identify approaches spanning multiple frameworks. Logical extensions such as FO-LTL could verify compliance properties over time, while encoding jurisdictional variation suggests applications in international governance. Overall, the metamodel provides a foundation for transparent data annotation and systematic evaluation of compliance status, in line with ISO recommendations~\cite{iso19944,iso38505,iso27001}.

\bibliographystyle{vancouver}
\bibliography{refs}

\begin{thebibliography}{10}

\bibitem{xacml}
{OASIS}.
\newblock eXtensible Access Control Markup Language (XACML) Version 3.0.
\newblock Organization for the Advancement of Structured Information Standards; 2013.
\newblock OASIS Standard.

\bibitem{nist-sp-800-95}
{National Institute of Standards and Technology}.
\newblock Guide to Secure Web Services.
\newblock NIST; 2007. SP 800-95.

\bibitem{nist-ir-7657}
{National Institute of Standards and Technology}.
\newblock A Report on the Privilege Management Workshop.
\newblock NIST; 2009. IR 7657.

\bibitem{hipaa-privacy}
Standards for Privacy of Individually Identifiable Health Information. {U.S. Code of Federal Regulations}; 2002.
\newblock 45 CFR Parts 160 and 164.

\bibitem{gdpr}
Regulation (EU) 2016/679 of the European Parliament and of the Council on the protection of natural persons with regard to the processing of personal data and on the free movement of such data (General Data Protection Regulation). {European Union}; 2016.
\newblock Official Journal L 119.

\bibitem{pandit2019dpv}
Pandit HJ, Polleres A, Bos B, Brennan R, Bruegger B, Ekaputra FJ, et~al.
\newblock Creating a {Vocabulary} for {Data} {Privacy}: {The} {First}-{Year} {Report} of {Data} {Privacy} {Vocabularies} and {Controls} {Community} {Group} ({DPVCG}).
\newblock In: Panetto H, Debruyne C, Hepp M, Lewis D, Ardagna CA, Meersman R, editors. On the {Move} to {Meaningful} {Internet} {Systems}: {OTM} 2019 {Conferences}. vol. 11877. Cham: Springer International Publishing; 2019. p. 714-30.
\newblock Series Title: Lecture Notes in Computer Science.

\bibitem{odrl-model-2018}
Iannella R, Villata S.
\newblock {ODRL} {Information} {Model} 2.2.
\newblock W3C; 2018.

\bibitem{odrl-vocab-2018}
Iannella R, McRoberts M, Villata S.
\newblock {ODRL} {Vocabulary} \& {Expression} 2.2.
\newblock W3C; 2018.

\bibitem{bader2020}
Bader S, Pullmann J, Mader C, Tramp S, Quix C, Müller A, et~al.
\newblock The {International} {Data} {Spaces} {Reference} {Architecture} {Model}.
\newblock International Data Spaces Association. 2020.
\newblock Version 3.0.

\bibitem{stalla_rossi_2021}
Stalla-Bourdillon S, Rossi A.
\newblock Aggregation, Synthesisation and Anonymisation: A Call for a Risk-based Assessment of Anonymisation Approaches.
\newblock In: Data Protection and Privacy. vol.~13. Bloomsbury Publishing Plc.; 2021. p. 111-43.

\bibitem{stalla_2025}
Stalla-Bourdillon S.
\newblock Identifiability, as a Data Risk: Is a Uniform Approach to Anonymisation About to Emerge in the EU?
\newblock European Journal of Risk Regulation. 2025 Jun.
\newblock Publisher Copyright: {\textcopyright} The Author(s), 2025. Published by Cambridge University Press.

\bibitem{srb}
{Court of Justice of the European Union}. EDPS v SRB; 2025.
\newblock ECLI:EU:C:2025:645.

\bibitem{ema_external_guidance_2016}
{External} {Guidance} on the {Implementation} of the {EMA} {Policy} on the {Publication} of {Clinical} {Data} for {Medicinal} {Products} for {Human} {Use} - Version 1.4.
\newblock {European Medicines Agency}; 2017. EMA/90915/2016.

\bibitem{ema_external_guidance_2025}
{External} {Guidance} on the {Implementation} of the {EMA} {Policy} on the {Publication} of {Clinical} {Data} for {Medicinal} {Products} for {Human} {Use} - Version 1.5.
\newblock {European Medicines Agency}; 2025. EMA/90915/2016.

\bibitem{wp29-anonymization}
{Article 29 Data Protection Working Party}.
\newblock Opinion 05/2014 on Anonymisation Techniques; 2014. WP216.

\bibitem{italian-thin-2023}
Provvedimento del 1 giugno 2023 - {Thin} {S.r.l.}. {Garante per la protezione dei dati personali}; 2023.
\newblock Provvedimento n. 226, doc. web n. 9913795.

\bibitem{lindenapotheke}
{Court of Justice of the European Union}. Lindenapotheke; 2024.
\newblock ECLI:EU:C:2024:846.

\bibitem{ema-policy0070}
{European Medicines Agency}.
\newblock {European} {Medicines} {Agency} {Policy} on {Publication} of {Clinical} {Data} for {Medicinal} {Products} for {Human} {Use}.
\newblock European Medicines Agency; 2019. EMA/240810/2013.

\bibitem{ocr-deid-guidance}
{Office for Civil Rights, U S  Department of Health and Human Services}.
\newblock Guidance Regarding Methods for De-identification of Protected Health Information in Accordance with the Health Insurance Portability and Accountability Act (HIPAA) Privacy Rule; 2012.

\bibitem{el-emam-2009}
El~Emam K, Dankar FK, Vaillancourt R, Roffey T, Lysyk M.
\newblock Evaluating the Risk of Re-identification of Patients from Hospital Prescription Records.
\newblock The Canadian Journal of Hospital Pharmacy. 2009;62(4):307-19.

\bibitem{wp29-personaldata}
{Article 29 Data Protection Working Party}.
\newblock Opinion 04/2007 on the concept of personal data; 2007. WP136.

\bibitem{oxigraph}
Pellissier~Tanon T. Oxigraph: A SPARQL graph database; 2025.
\newblock Available from: \url{https://github.com/oxigraph/oxigraph}.

\bibitem{parajudica}
Moreau L, Rossi A, Stalla-Bourdillon S. Parajudica: An RDF-Based Reasoner for Multi-Framework Compliance; 2025.
\newblock GitHub repository containing the RDF/SPARQL implementation and example frameworks.
\newblock \url{https://github.com/alfredr/parajudica}.

\bibitem{jena}
{The Apache Software Foundation}. Apache Jena: A free and open source Java framework for building Semantic Web and Linked Data applications; 2011--2024.
\newblock \url{https://jena.apache.org/}.

\bibitem{iso27560}
{International Organization for Standardization}.
\newblock Privacy technologies: Consent record information structure.
\newblock Geneva, CH: International Organization for Standardization; 2023.
\newblock ISO/IEC TS 27560:2023. Accessed: 01.09.2023.

\bibitem{legalruleml}
Athan T, Governatori G, Palmirani M, Paschke A, Wyner A.
\newblock LegalRuleML Core Specification Version 1.0.
\newblock OASIS; 2021.
\newblock Committee Specification 02.

\bibitem{prov-o}
Lebo T, Sahoo S, McGuinness D.
\newblock {PROV-O}: The {PROV} {Ontology}.
\newblock W3C; 2013.
\newblock W3C Recommendation.

\bibitem{dcat}
Albertoni R, Browning D, Cox S, Gonzalez~Beltran A, Perego A, Winstanley P.
\newblock Data {Catalog} {Vocabulary} ({DCAT}) - Version 2.
\newblock W3C; 2020.
\newblock W3C Recommendation.

\bibitem{fibo}
{Enterprise Data Management Council}. Financial {Industry} {Business} {Ontology} ({FIBO}). Enterprise Data Management Council; 2023.
\newblock EDM Council Standards.

\bibitem{gist}
McComb D, Smith M. Gist: {Upper} {Ontology} for {Enterprise} {Modeling}. Semantic Arts; 2022.

\bibitem{iso19944}
{International Organization for Standardization}.
\newblock Cloud computing and distributed platforms: Data flow, data categories and data use, Part 1: Fundamentals.
\newblock Geneva, CH: International Organization for Standardization; 2020.
\newblock ISO/IEC 19944-1:2020. Accessed: 01.09.2023.

\bibitem{iso38505}
{International Organization for Standardization}.
\newblock Information technology: Governance of data, Part 3: Guidelines for data classification.
\newblock Geneva, CH: International Organization for Standardization; 2021.
\newblock ISO/IEC TS 38505-3:2021. Accessed: 01.09.2023.

\bibitem{iso27001}
{International Organization for Standardization}.
\newblock Information security management systems.
\newblock Geneva, CH: International Organization for Standardization; 2023.
\newblock ISO/IEC 27001:2023. Accessed: 01.09.2023.

\end{thebibliography}

\end{document}